\documentstyle[proceed,graphicx]{article}

\long\def\comment#1{}

\long\def\symbolfootnote[#1]#2{\begingroup
\def\thefootnote{\fnsymbol{footnote}}\footnote[#1]{#2}\endgroup}

\title{Big-Five Personality Prediction\\
Based on User Behaviors at Social Network Sites}

\author{ {\bf Shuotian Bai, Tingshao Zhu\thanks{Corresponding author: tszhu\char'100 gucas.ac.cn}} \\
Graduate University of Chinese Academy of Sciences \\
Beijing 100190, China\\
\And
{\bf Li Cheng}  \\
Bioinformatics Institute, A*STAR \\
Singapore \\
}

\begin{document}

\maketitle

\begin{abstract}

Many customer services are already available at Social Network Sites (SNSs), including user recommendation and media interaction, to name a few. There are strong desires to provide online users more dedicated and personalized services that fit into individual's need, usually strongly depending on the inner personalities of the user. However, little has been done to conduct proper psychological analysis, crucial for explaining the user's outer behaviors from their inner personality. In this paper, we propose an approach that intends to facilitate this line of research by directly predicting the so called Big-Five Personality from user's SNS behaviors. Comparing to the conventional inventory-based psychological analysis, we demonstrate via experimental studies that users' personalities can be predicted with reasonable precision based on their online behaviors. Except for proving some former behavior-personality correlation results, our experiments show that extraversion is positively related to one's status republishing proportion and neuroticism is positively related to the proportion of one's angry blogs (blogs making people angry).

\end{abstract}

\section{INTRODUCTION}

Personality is the particular combination of emotional, attitudinal, and behavioral response patterns of an individual in psychological definition (Wikipidia, 2012). According to the classic Big Five Personality traits theory, personality can be divided into five different dimensions which are agreeableness, conscientiousness, extraversion, neuroticism and openness for most cases (Lounsbury, 2006). Agreeableness refers to being helpful, cooperative, and sympathetic towards others. Conscientiousness is determined by being disciplined, organized, and achievement-oriented. Extraversion is displayed through a higher degree of sociability, assertiveness, and talkativeness. Neuroticism refers to degree of emotional stability, impulse control, and anxiety. Finally, openness is reflected in a strong intellectual curiosity and a preference for novelty and variety. (Funder, 2001)

To analyze individual personality is extremely important for many researches. Canada Peer Counseling Center (Chen, 1998) considers that for most people they have investigated, recommendations from companion volunteers with same world view as being the most effective. It means people with same personality tend to attract each other. As a result, analysis of different personality features can be the basis for building characterized service. For example, an extraversive user may have a higher level of online activity which is more likely to use recommendation system to make new friends with strangers (McElroy, 2012).

Analyzing outer behaviors is the principle of inner personality analysis since behavior is the manifestation of personality.  In psychological researches, most traditional personality analyzing experiments are based on self-reported inventory. However, psychological experiment has its own bottleneck. When experiment participants upload the self-report data, they could have reflected self-views rather than actual behavior. Other data collection methods such as observable information profile cost a lot of manual resources and are not desirable for large scale of dataset collection. At the same time, most personality researches can only build the covariation relation between behavior and personality instead of a quantitative personality prediction.

Facing these disadvantages, we propose an automatic and objective personality prediction system based on user's behaviors on Social Network Sites (SNSs). Online Social Networking Sites (SNSs) like Facebook and RenRen (Renren, 2012) have a quick development during the recent decade. They have already been a part of people's life and an extension of real nature. According to the Chinese social e-commerce Report from IResearch, Chinese SNSs have totally 370 million registered users in the year 2011 which gets an increase of 17.6\% compared with the previous year. It is predicted that the users count will jump to 510 million in 2014 (IResearch, 2011). RenRen, a Chinese version Facebook, has the highest Market share in Chinese SNSs (Baidu, 2009). RenRen provides a wide range of functions for information exchanging where people can keep connections with each other such as blogging, status, and photo/video-sharing (Boyd, 2007).

Online SNS behaviors and real world behaviors have a lot in common (Lounsbury, 2006). Self-report and interactive behaviors are all supported in SNS. Therefore, many experts tend to do research on this field. Techniques on computer science such as Information Retrieval (IR) and Recommendation System are helpful to solve many problems using keyword-resource matching and collaborative filtering methods. However, along with the social network functions improvement, it is unavoidable to consider highly for the user experience since the user demand is increasing. The more characterized systems that connect the network behavior based personal preference and online resources are welcomed.

Since computer science needs psychological characterized service and psychology needs automatic computation, we come up with the idea of building the relation between personality and online behavior in Renren which uses an automatic computation to predict user's personality attributes. With our model, user's big-five personality can be predicted based on her SNS behavior. In the following section 2, we will show some related work on both computer science and psychology. Then we will explain our researching methods in section 3. In section 4, we will show our experiment results. Finally in section 5, we will make a conclusion and discuss our further work.

\section{RELATED WORK}

Previous researches on SNS mostly focus on topological characteristics (Kwak, 2007), web community mining(Kevin, 2010) and so on. From these meaningful results, virtual world is a facsimile version of the nature society which follows most sociological principles such as Six Degrees of Separation and Rule of 150 (Yaguang, 2009). It is also found that online users tend to join with each other to form some small communities. Meanwhile, the growing user demand in SNS world triggers the taking off for techniques of characterized recommendation (Jie, 2011) and information retrieval (Christopher, 2010) recent year. Junco Reynol (Reynol, 2011) researched on relationship between Facebook use and student engagement and found that Facebook use was negatively predictive of engagement scale score and positively predictive of time spent on SNS. However, these works were based on user's statistic information such as common friend count, familiar shared resources, time spent on SNS or information checked frequency which considers user's SNS usage instead of her inner preferences and personality.

Personality is one of the hottest topics in Psychology. According to Big Five personality traits theory, personality can be divided into five different dimensions which are openness, conscientiousness, extraversion, agreeableness and neuroticism. Berkeley Personality Lab (Berkeley, 2012), focusing on personality, self-perception, and individual differences in emotion regulation, designed a Big Five Inventory which is wildly used around world. It contains 44 questions with high validity and reliability and can give back a quantized personality score with five dimensions.

Until now, researches that combine personality and SNS together have a few bases (Shaoqi, 2011). Emily S. Orr discussed the influence of shyness on the use of SNS in undergraduate samples in 2009. He discovered that shyness was significantly positively correlated with the time spent on SNS and negatively correlated with the number of "friends" (Sisic, 2009). Meanwhile, Teresa Correa analyzed the intersection of users' personality and social media (Correa, 2010) and found that openness and extraversion had positive relation to using experience of social media while neuroticism was a negative predictor. However, these works could only give the association relation between personality and behavior instead of a quantization of personality metrics.

Samuel D. Gosling (Gosling, 2011) experimented on the manifestations of personality in SNS. In this research, a mapping between personality and SNS behavior is announced. They examined the personality with self-reported Facebook usage and observable profile information and finally gave the correlation factor between personality and online behavior. They designed 11 features, friends count, weekly usage and 9 other functions using frequency. However, their features are all based on statistical characteristics without any inner properties of user. The data collections are based on self-reported usage and observable profile information which will need a large amount of manual operation. Therefore, experiment objectivity will get a discount.

Generally speaking, most researches on personality used only psychological method. No matter self-report or observable information profile, they are all not efficient for large large-scale data acquisition. At the same time, the features they used are only from SNS statistic frequency usage. It would be better if some emotion-related features (e.g. blog emotion, anger or happiness) could be added. The association mode between personality and SNS behavior could only give the correlation factor instead of predicting personality. Although these factors can describe the relationships between personality and behavior, they can't accurately quantify personality for an arbitrary testing sample. Since psychology and computer science have their own advantages as well as disadvantages, we try to cross these two subjects and build a predictor system that can qualify user's big five personality based on SNS usage and preferences.

\section{METHODS}

In our work, we try to build a personality computation and prediction model based on user's online SNS usage. We choose the most wildly used Chinese SNS Renren as our experiment platform. In this part, we will solve the following problems:\\
\emph{How could we collect a large amount of dataset objectively and efficiently?}\\
\emph{How could we design the features that distinguishing different personality preference?}\\
\emph{How could we build the personality computation and prediction model?}\\

\subsection{DATA COLLECTION}

Renren has opened a lot of APIs for third-side application design (Platform, 2012). These third-side applications can be divided into three classes, Web Access Connections, Web/Wap Applications and Mobile Client Applications. We have already developed an online mental illness treatment website Dao (Dao, 2012). We then make it a web access connection to Renren which allows Renren users login in DAO through her/his Renren account. When our experiment participants login DAO, user authorization can be achieved. Then we can call for APIs of Renren to collect user historical behavior data and save it into our local database.

In order to get the labeled data, an inventory is asked for finishing by each participant. The Big Five Inventory (BFI), designed by Berkeley Personality Lab, is a self-report inventory to measure the Big-Five personality dimensions. It is quite brief for a multidimensional personality inventory (44 questions in all) and including short phrases with relatively accessible vocabulary (Berkeley, 2012). After finishing this inventory, a personality result vector with five dimensions can be saved as data labels. Therefore, with these computer application techniques, building a labeled dataset with high efficiency can be easily achieved.

\subsection{FEATURE DESIGN}

The initial data we collect could not be used directly, so we design 41 features based on BFI and some previous work in this field to describe user behavior. The features can be divided into 5 groups each of which is listed in the table~\ref{Features} below, where T stands for time, E stands for emotion.

\begin{table}[h]
\caption{Features Design}
\label{Features}
\begin{center}
\begin{tabular}{lll}
\multicolumn{1}{c}{\bf FEATURE GROUP}  &\multicolumn{1}{c}{\bf COUNT}  \\
\hline \\
   Basic Info. &5 \\
   SNS Usage	&28	\\
   T-Related Usage	&3	\\
   E-Related Usage	&2	\\
   T\&E-Related Usage	&3	\\
\end{tabular}
\end{center}
\end{table}

Features of basic information and SNS usage experiences have already been used by a lot of previous work including the researches we listed in section 2. These features contain user's gender, age, hometown and blog usage frequency, resource uploading frequency and so on. The
time-related features include the features that correlated with the recent psychological state such as status or blog publishing count during recent one month. The emotion-related features mean the features that related with the emotion distribution (angry, funny, surprised and moving) of the user such as the top emotion count of all her blog. We would like to find out the emotion distribution of the user and select the top emotion count. These features stand for the user's emotion preferences and have a strong relation with her personality. The final feature class time\&emotion-related features take both time and emotion into account which means the recent emotion tendency of the user such as the status emotion of the newest status and its emotion length. Emotion length means the time sustained of the recent emotion.

Take a look at the emotion-related features, they need an emotion predictor. That's our previous work last year. Using Naive Bayesian method, the system is a combination of text classification with emotion dictionary. The key idea is increasing the weight of emotion token and decreasing non-emotion token while training the model. We have already tested on a large scale of text content and get a high accuracy and recall rate over 80\%. The result is to classify an article into different emotion (angry, funny, surprised and moving) according to its content.

\subsection{SYSTEM DESIGN}

Our system comes from a combination of machine learning from Computer Science and Big Five Inventory from Psychology shown in figure below.

\begin{figure}[h]
\label{Flowingchart}
\vspace{1in}
\includegraphics[width=3.2in]{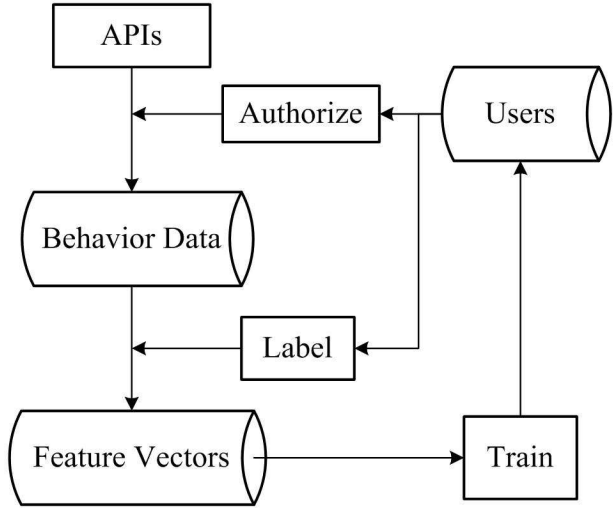}
\caption{System Flowing Chart}
\end{figure}

We call for the Application Program Interfaces (APIs) of Renren and collect users' online behaviors containing user basic profile, basic functions usage frequency and blog/status text content. In order to get all the information, experiment users need to give us authorization for using APIs. We develop a Web Access Connection to Renren.com that allows users to authorize on Dao, an experiment platform we design. Then users need to finish Big Five Inventory and label her/his behavior data with the inventory results. Finally, using data mining techniques, train a prediction model based on feature vectors.

\section{EXPERIMENT}

\subsection{SAMPLES}

We have developed an experiment platform Dao in which participants can login with her/his Renren account. Participants can do the experiment everywhere through networking, even at their dormitories. In order to keep the quality of training samples, a testing fee is given to everyone after they finish the inventory. We advertise our experiment around Graduate University of Chinese Academy of Sciences (GUCAS) and get 335 participants in January and February 2012. Each of them is shown with the informed consent telling them that we will collect their Renren usage data. All the participants are friend or friend's friend of us from China with average $23.833$ years old. However, the participants need to carefully finish the inventory, be active user in Renren with friends count over 100, status count over 50 and blog count over 10. Finally with these principles, we select 209 of them as our training dataset with 72 females and 137 males.

\subsection{PRE-PROCESSING}

\begin{figure}[h]
\includegraphics[width=3.2in]{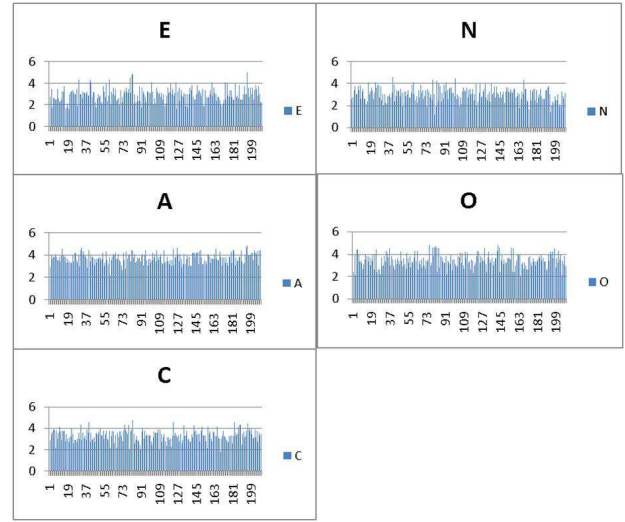}
\caption{Personality Score Distribution} \label{personalityscore}
\end{figure}

After collecting the behavior data of all the legal participants, we need to label each sample with five personality dimensions score. However, the scores are continuous value ranged from one to five that can't be used directly for classification shown in figure~\ref{personalityscore} where the horizontal axis stands for participant IDs, the vertical axis stands for her/his one dimension score, A stands for agreeableness, C stands for conscientiousness, E stands for extraversion, N stands for neuroticism and O stands for openness.

In order to train our system using classification methods in machine learning, we need to do the discretization on the initial scores and use the discrete values as data labels. The discretization functions we used are shown below:
\begin{center}
$\alpha = E(x) - \sigma(x)$, \\
$\beta = E(x) + \sigma(x)$,
\end{center}
For each dimensions, it means that we separate the label scores into three classes, low-score group 1 to $\alpha$, middle-score group $\alpha$ to $\beta$ and high-score group $\beta$ to 5, where E(x) is the mean value of personality score for dimension x, $\sigma(x)$ is the Standard variation of dimension x and $x$ is choosen from E,A,C,N,O, the five dimensions for personality. Therefore, we change the label distribution from figure~\ref{personalityscore} to table~\ref{Discretization} shown below. The three numbers in column sample count means the sample count in each group. For example dimension E, there are 62 samples in low-score group $1\sim2.31(\alpha)$, 92 samples in middle-score group $2.31\sim3.59(\beta)$ and 55 samples in high-score group $3.59\sim5$.

\begin{table}[h]
\caption{Discretization}
\label{Discretization}
\begin{center}
\begin{tabular}{llll}
\multicolumn{1}{c}{\bf PERSONALITY}  &\multicolumn{1}{c}{\bf $E(x)$}&\multicolumn{1}{c}{\bf $\sigma(x)$}&\multicolumn{1}{c}{\bf COUNT}  \\
\hline \\
   E    &2.95   &0.64   &62,92,55\\
   A	&3.71   &0.47   &69,67,73\\
   C	&3.29   &0.55   &72,78,59\\
   N	&3.02   &0.61   &54,89,66\\
   O	&3.39   &0.61   &56,82,71\\
\end{tabular}
\end{center}
\end{table}

\subsection{MODEL TRAINING AND TESTING}

Until now, we have changed the whole work into a classification problem. We test the dataset on many classification algorithms such as Naive Bayesion (NB), Support Vector Machine (SVM), Decision Tree and so on. We find out that C4.5 Decision Tree (Quinlan, 1993) can get the best results. Using 10-fold cross validation, the results (precision, recall and F-value) of three-class classification problem for five personality dimensions are shown in table~\ref{tab:Three-Class}.

\begin{table}[h]
\caption{Three-Class Classification}
\label{tab:Three-Class}
\begin{center}
\begin{tabular}{llll}
\multicolumn{1}{c}{\bf DIMENSION}  &\multicolumn{1}{c}{\bf P}&\multicolumn{1}{c}{\bf R}&\multicolumn{1}{c}{\bf F-VALUE}  \\
\hline \\
A&	0.725&	0.722&	0.723\\
N&	0.713&	0.708&	0.710\\
C&	0.702&	0.703&	0.701\\
E&	0.718&	0.718&	0.717\\
O&	0.697&	0.694&	0.695\\
\end{tabular}
\end{center}
\end{table}

We also consider removing the middle-score group and experiment on the low and high groups for each dimension, that are 1 to $\alpha$ and $\beta$ to 5 intervals. In this testing, we delete the "middle personality" samples and consider the two extreme personality cases only. That changes the work into a two-class classification problem. Still using C4.5 Decision Tree, results are shown in table~\ref{tab:Two-Class}. Since the problem is simplified to a two-class classification problem, all the result quotas get a rise.

\begin{table}[h]
\caption{Two-Class Classification}
\label{tab:Two-Class}
\begin{center}
\begin{tabular}{llll}
\multicolumn{1}{c}{\bf DIMENSION}  &\multicolumn{1}{c}{\bf P}&\multicolumn{1}{c}{\bf R}&\multicolumn{1}{c}{\bf F-VALUE}  \\
\hline \\
A&	0.697&	0.697&	0.697\\
N&	0.749&	0.750&	0.749\\
C&	0.825&	0.824&	0.824\\
E&	0.839&	0.838&	0.838\\
O&	0.811&	0.811&	0.811\\
\end{tabular}
\end{center}
\end{table}

We show the root part of the decision tree of dimension A in figure 3 since the whole tree is too long to display. Feature zidou, the root node, means the virtual money of the user's account. The following 2nd root node selfcommentproportion means the proportion of the comment from herself. Blogemocticon means the count of emoticon used in the user's blogs.  The 3rd root node recentstatustopemotionratio means the ratio of the majority emotion count in recent one month. For example, there are 10 pieces of status in the recent month of the user and 6 of them will make the reader angry, 4 of them will make reader happy. Recentstatustopemotionratio will be set 0.6. BlogIYouIt means which person the user likes to use, I, you or it (including "he", "she" and "they").

\begin{figure}[h]
\vspace{1in}
\includegraphics[width=3.2in]{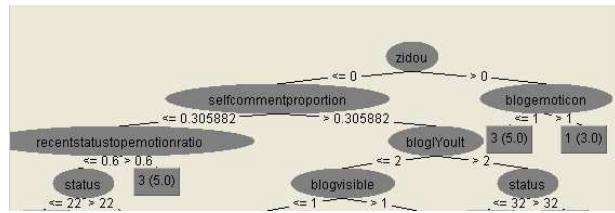}
\caption{3-Class Decision Tree For A}
\end{figure}

\section{DISCUSSION}

From the above results, it is easy to see that different attribution has different weight and different dimension has different high-weight attribution. That proves the difference of these five personality dimension and their behavior performances which are the strong evidence of the correlation of behavior and personality. We will give a discussion on the five dimensions from the view of the decision trees. As we all know, the features near the root nodes have a strong classifying contribution.

\subsection{RESULTS ANALYSIS}

The classifying algorithm we use is C4.5 decision tree which uses Gain Ratio (Han and Kamber, 2008) to extract features. The purpose of feature extraction is to find a splitting principle that can best predict the results.  Gain ratio, a normalized information gain, stands for classification contribution of features. Calling for C4.5 decision tree, the algorithm calculates the gain ratio of each feature and set the feature with highest gain ratio as the root node. Repeating this operation, we can get a tree with high-gain-ratio features above and low-gain-ratio features below. Therefore, we list the root nodes (highest gain ratio features) and second root nodes (2nd gain ratio features) for five dimensions in table~\ref{Strong}, where p(x) means the proportion of variable x.

\begin{table}[h]
\caption{Strong Contribution Features}
\label{Strong}
\begin{center}
\begin{tabular}{lll}
\multicolumn{1}{c}{\bf D.}  &\multicolumn{1}{c}{\bf LEVEL}&\multicolumn{1}{c}{\bf FEATURES}\\
\hline \\
   A&   Root&	zidou\\
    &   2nd Root&	p(selfcomment), blogemoticon\\
   C&   Root&	age\\
    &   2nd Root&	p(friendcomment), guestbook\\
   E&   Root&	friend\\
    &   2nd Root&	blogemoticon, zzstatus\\
   N&  Root&	friend\\
    &   2nd Root&	usage, p(angryblog)\\
   O&   Root&	friend\\
    &   2nd Root&	usage, recentstatus\\
\end{tabular}
\end{center}
\end{table}

From the table, we can draw a lot of interesting and meaningful conclusions. Dimension agreeableness refers to being helpful, cooperative, and sympathetic towards others. People with high scores in agreeableness tend to send more blogs or emails (YANG, 2007). In SNS, this is reflected on interaction between users. For user having more virtual money (zidou), they are more likely to buy virtual gifts for others. A person with a high score on agreeableness tends to be more active on chatting online even others will discard her message. Therefore, their self-comment proportion is relatively high. Also in order to get the attention of others, they may be more likely to use emoticons (blogemoticon) in their blogs.

Dimension conscientiousness is judged by being disciplined, organized, and achievement-oriented. People using guestbook are most likely to call for some help from others such as asking for a location or an email-address. People having a high score in this dimension tend to be helpful for others and will use guestbook more frequently.

Dimension extraversion is displayed through a higher degree of sociability, assertiveness, and talkativeness. It is easy to find that people with more friends(friend) is more likely to be extraversive. For an extraversive person, she may tend to use emoticon in her blogs to show her character. She has many friends and is happy to talk with others even to republish (zzstatus) others' statuses.

Dimension neuroticism refers to degree of emotional stability, impulse control, and anxiety. Clearly, people that have a high score in this dimension tend to be easily angry for other things. Therefore, their blogs may have a high proportion of making readers angry. That is positively related to the proportion of angry blogs (angryblogproportion).

Dimension openness is reflected in a strong intellectual curiosity and a preference for novelty and variety. People that are curious with others tend to make many new friends and have a high SNS usage experiences(usage). Their statuses tend to be updated frequently which means their recent status count (recentstatus) is relatively high. As in Correa's work (Correa, 2010), openness is positively correlated with SNS usage which holds our results.

\subsection{CONCLUSION AND FUTURE RESEARCH}

The automatic personality predicting can open a new window not only for computer science but also for psychology. SNS servicer can recommend resources based on user's personality in the future. For outgoing users, he may prefer international news and like to make friends with others, which will be the guides for networking service suppliers. Also, an objective data selection strategy is given to psychological experiments which will increase their quality levels and confidence degrees. This system can also be used into online mental illness treatment. For extroverted patients, she/he may easily explain her/his illness to the doctor without lies while introverted patient may speak a little about her/his illness which calls for doctor's patience.

We will continue our work on this Cross discipline topic. To make the whole system better, we may consider the correlation of these five personality dimensions, since these five dimensions are not absolutely orthogonal. We may use multi-task learning techniques to fix our training algorithm. At the same time, the consistency of online behavior and offline behavior is another interest for us. Although the networking developing tendency is to build a virtual world quite same as nature world, there are still some differences on user behavior between online and offline. Most online services are based on the real-name system, but it is still not face-to-face. Users do not need to consider cases of losing face which in nature world consider highly. We believe there must be but a small difference between online and offline behavior.

\subsubsection*{Acknowledgements}

The authors gratefully acknowledges the generous
support from NSFC(61070115), Research-Education Joint
Project(110700EA02) from Chinese Academy of Sciences,
and the Scientific Research Foundation for the Returned
Overseas Chinese Scholars(Y01Z0311A9), State Education Ministry.

\subsubsection*{References}

{\it Chinese social e-commerce Report}. http://www.
iresearch.cn/, (2011).

{\it Wikipidia}.http://en.wikipedia.org/,(2012).

{\it Renren}.http://www.renren.com/, (2012).

{\it Berkeley}.http://www.ocf.berkeley.edu/\~johnlab/index.htm, (2012).

{\it Renren open application platform}. http://dev.renren.com, (2012).

{\it Dao}. http://dao.gucas.ac.cn, (2012).

Ross Quinlan. C4.5: {\it Programs for Machine Learning}. Morgan Kaufmann Publishers, San Mateo, CA. (1993).

Jiawei Han, Micheline Kamber. {\it Data Mining Concepts and Techniques, second edition}. Elsevier (Singapore) Pte Ltd. 2008.

Han S. Kwak H. Moon S. Ahn, Yong-Yeol and
H Jeong. {\it Analysis of topological characteristics of
huge online social networking services}. WWW¡¯07:
Proceedings of the 16th international conference on
World Wide Web, 835-844, (2007).

Danah Boyd and Nicole. Ellison. {\it Social network
sites: Definition, history, and scholarship}. JCMC,, 13,
(2007).

Guohai Chen. {\it The peer help centre in Canadian colleges
and its inspiration}. Researches on Higher Education,
(1998).

Christopher DManning. {\it Introduction to Information
Retrieval}. People¡¯s Posts and Telecommunications
Press, (2010).

Mia Sisic B.A. Craig Ross M.A. Mary G. Simmering
M.A. Jaime M. Arseneault M.A. Emily S. Orr, M.A.
and Ph.D. R. Robert Orr. {\it The influence of shyness on
the use of facebook in an undergraduate sample}. Cyberpsychology
\& Behavior, Volume 12,:337–
340, Number 3, (2009).

David C. Funder. {\it Personality}. Annu. Rev. Psychol.,
52:197-221, (2001).

Xu Jie. {\it Sequencing algorithm based on the social network
real-time search engine}. Science Technology and
Engineering, page 1671-1815, (2011).

James C. McElroy Kelly Moore. {\it The influence of personality
on facebook usage, wall postings, and regret}.
Computers in Human Behavior, 28:267-274, (2012).

Baidu Lib. {\it Chinese social networking sites developmental
reports}. (2009).

Ma Shaoqi; Jiao Can; Zhang Minqiang.
{\it Application of social network analysis in psychology}.
Advances in Psychological Science, Vol. 19, No.
5,:755-764, (2011).

Kevin Macon Mason A. Porter Peter J. Mucha,
Thomas Richardson and Jukka-Pekka Onnela.
{\it Community structure in time-dependent, multiscale,
and multiplex network}. Science 14,
:876-878.(2010).

Junco Reynol. {\it The relationship between frequency
of facebook use, participation in facebook activities,
and student engagement}. Computers \& Education,
doi:10.1016/j.compedu.2011.08.004, (2011).

JohnW. Lounsbury Richard N. Landers. {\it An investigation
of big five and narrow personality traits in relation
to internet usage}. Computers in Human Behavior 22,
page 283C293, (2006).

1 Adam A Augustine M.S. 2 Simine Vazire Ph.D. 2
Nicholas Holtzman M.A. 2 Samuel D. Gosling, Ph.D.
and B.S.1 Sam Gaddis. {\it Manifestations of personality
in online social networks: Self-reported facebookrelated
behaviors and observable profile information}.
Cyberpsychology, Behavior, and Social
Networking, 14:483-488, Number 9, (2011).

SHI Yaguang. {\it Criticism of the category of world literature
in chinese library classification}. Library Tribune,
29:220-223, (2009).

Teresa Correa, Amber Willard Hinsley, and Homero Gil de Ziga.
{\it Who interacts on the web?: The intersection of users' personality
and social media use}. Computers in Human Behavior, 26:247-253,
(2010).

Yang Yang and Pi Lei. {\it The relationship between adolescents extra versionp agreeableness
, internet service preference , and internet addiction}. Psychological Development and Education, pages 42-48,
(2007).

\end{document}